\documentclass[journal=jacsat,manuscript=article]{achemso}

\usepackage{chemformula} 
\usepackage[T1]{fontenc} 
\usepackage{xcolor}
\usepackage{soul}

\usepackage{subfigure}

\author{Menno Demmenie}
\affiliation{Institute of Physics, University of Amsterdam, Science Park 904, 1098 XH Amsterdam, the Netherlands}
\alsoaffiliation{Van ’t Hoff Institute for Molecular Sciences, University of Amsterdam, Science Park 904, 1098XH Amsterdam, The Netherlands}
\email{m.demmenie@uva.nl}
\author{Paul Kolpakov}
\affiliation{Institute of Physics, University of Amsterdam, Science Park 904, 1098 XH Amsterdam, the Netherlands}
\author{Yuki Nagata}
\affiliation{Max Planck Institute for Polymer Research, Ackermannweg 10, 55128 Mainz, Germany}
\author{Sander Woutersen}
\affiliation{Van ’t Hoff Institute for Molecular Sciences, University of Amsterdam, Science Park 904, 1098XH Amsterdam, The Netherlands}
\author{Daniel Bonn}
\affiliation{Institute of Physics, University of Amsterdam, Science Park 904, 1098 XH Amsterdam, the Netherlands}

\title[An \textsf{achemso} demo]
  {Self-Healing Behavior of Ice}

\abbreviations{IR,NMR,UV}
\keywords{American Chemical Society, \LaTeX}

\begin{document}

\begin{tocentry}

\includegraphics[scale=0.382]{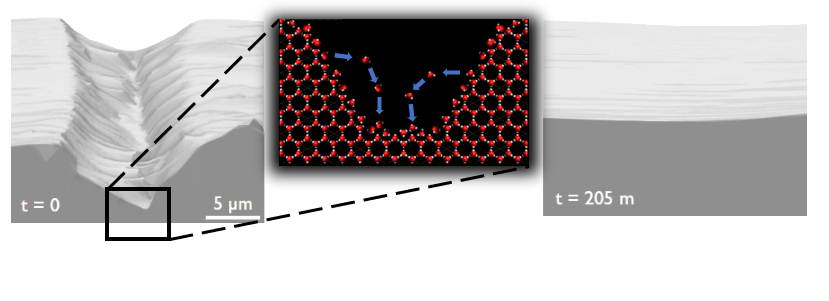}

\end{tocentry}

\begin{abstract}
    We show that the surface of ice is self-healing: micrometer deep scratches in the ice surface spontaneously disappear by relaxation on a time scale of roughly an hour. Following the dynamics and comparing it to different mass transfer mechanisms, we find that sublimation from and condensation onto the ice surface is the dominant self-healing mechanism. The self-healing kinetics shows a strong temperature dependence, following an Arrhenius behavior with an activation energy of $\Delta E = 58.6 \pm 4.6$ kJ/mole, agreeing with the proposed sublimation mechanism, and at odds with surface diffusion or fluid flow or evaporation-condensation from a quasi-liquid layer. 
\end{abstract}

    Ice is one of the most actively studied solids\cite{shultz2017ice,hudleston2015structures,blackford2007sintering,li2012freeze,bartels2014review,petrovic2003review,nagata2019surface,schutzius2015physics,russo2014new,korolev2020review,gelman2018quasi}, and many of its physical properties are still poorly understood. In particular, the structure and dynamics of the topmost layer of water molecules in ice has been subject of intense debate for more than 150 years. Already in 1860, Faraday observed that ice cubes sinter together, and he concluded that there is always a liquid layer present on the ice surface, even at atmospheric temperatures far below the melting temperature \cite{faraday1860note, thomson1860recent}. It took more than 100 years until detailed measurements of the speed of sintering finally ruled out the idea that the flow of a liquid layer was at the origin of the sintering dynamics \cite{blakely1961surface,king1962theory,kingery1960regelation, van2013experimental, chen2010structural, nie2009observation}. Explanations of the mass transfer phenomenon underlying ice sintering included highly mobile surface molecules undergoing surface diffusion \cite{kingery1960regelation}, bulk lattice motion \cite{kuroiwa1961study}, and condensation from the vapor phase \cite{chen2013surface,gundlach2018sintering, molaro2019microstructural}. More recent research suggests that the outermost molecular layer of an ice crystal is disordered\cite{pickering2018grand,kahan2007spectroscopic,qiu2018so,slater2019surface,murata2016thermodynamic}. However, this 1 to 2 molecules thick layer cannot simply be considered as a liquid since it exhibits viscoelastic properties  \cite{Canale2019,louden2018ice}. \color{black} General crystal growth theory is insufficient to describe the diffusion limited dynamics of ice crystals since it does not account for the ambiguous disordered interface, cooperative intermolecular hydrogen bonding and the degree of supercooling. Moreover, no model has proposed that completely describes the unique growth behavior of ice thus far.\cite{libbrecht2017physical,kelly2014physical,gravner2008modeling,zhao2020review} \color{black} 
    
    To shed new light on the complex molecular dynamics of the surface of ice we investigate the temporal evolution of a scratch made in a pristine surface of ice with sub-micrometer precision, under precisely controlled experimental conditions. We find that the scratch heals spontaneously over time, and that eventually the ice surface becomes completely smooth again.
    By comparing the data quantitatively to models for the different proposed healing mechanisms, we conclude that self healing of ice occurs by the detachment and reattachment of surface molecules. Since the transport of water molecules in the ambient phase is limited by diffusion, this process is dominated by local sublimation from and condensation onto the surface\cite{jambon2018singular}. The obtained activation energy corresponds to the known value for sublimation, which is significantly higher than the energy barrier for liquid evaporation. This settles the longstanding debate on the origin of the sintering dynamics of ice, and explains why this phenomenon is unique to ice with its high vapor pressure. We perform the same measurements on a metal exhibiting surface premelting and a crystalline mineral: both do not show self healing.

\begin{figure*}
  \centering
  \includegraphics[scale=1.46]{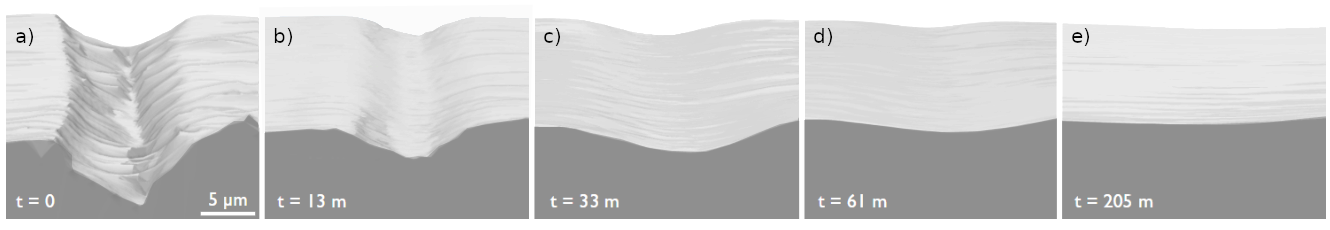}
  \caption{\label{Figure} Evolution of a scratch in ice (initial depth $\sim$2.5$\mu$m) healing in time under controlled conditions, with constant ice temperature of 247 K and vapor pressure at equilibrium.}
  \label{Fig:3DPhotographsInTime}
\end{figure*}

Measurements were carried out with a confocal profilometer (Keyence VK-X1100), with a lateral resolution of 212 nm and a vertical resolution of 12 nm\cite{keyence2018manual}, in a temperature and humidity controlled chamber. The humidity was regulated by an inflow of dry nitrogen into the chamber and monitored by a Thermo/Hygrometer (Testo 645, error of $0.1 \%$ relative humidity). Furthermore, to reach low humidities, a controlled flow of liquid nitrogen (Norhof microdosing LN2, 900 series) through a copper element acted as a cold trap to remove remaining water vapor by condensation. Hence, a theoretical equilibrium between the vapor pressure of the flat ice surface and the air could be achieved, which was calculated with parameters of Murphy and Koop\cite{murphy2005review}. When the measured humidity deviated from the theoretical equilibrium by a larger value than the error margin of the hygrometer, the inflow of nitrogen was adjusted. Apart from the healing of scratches no sign of sublimation nor condensation was observed on the horizontal surfaces, confirming the stability of the equilibrium in the chamber.

The ice layers were formed by cooling down 3 mL ultra-purified water from a Milli-Q system on a copper plate (560$\times$380$\times$40 mm). Cooling was done by a Peltier element in direct contact (using thermal paste) with the copper plate to ensure an isothermal and homogeneous layer of ice. The induced heat on the opposite side of the Peltier element was extracted from the system by flow from a temperature bath. Ice temperatures were in the range from 243.0 K to 272.6 K (measured by a Voltcraft PL-125-T2USB VS temperature probe). Micron-sized scratches were manually created with a sharp razor blade (Derby extra Paslanmaz \c{C}elik), and positioned in such a way that the measured area of 212 by 283 $\mu$m contained only one defined crystal orientation, so that grain-boundary dynamics could be excluded. Since different grains do not exhibit wide variations in molecular organisation at the surface, we were enabled to collect an ensemble of measurements performed under similar experimental conditions, as the effective diffusion coefficient is expected to be similar\cite{huang2013size}. 
The profile of the ice was regularly monitored by a 50$\times$ magnification Plan Apo objective (NA 0.95, WD 0.35 mm, 404 nm wavelength reflection) for the period of self-healing. For a detailed 3D model of the setup, see Supplementary Fig.~6. 

\begin{figure}[h]
  \centering
  \includegraphics[scale=0.4]{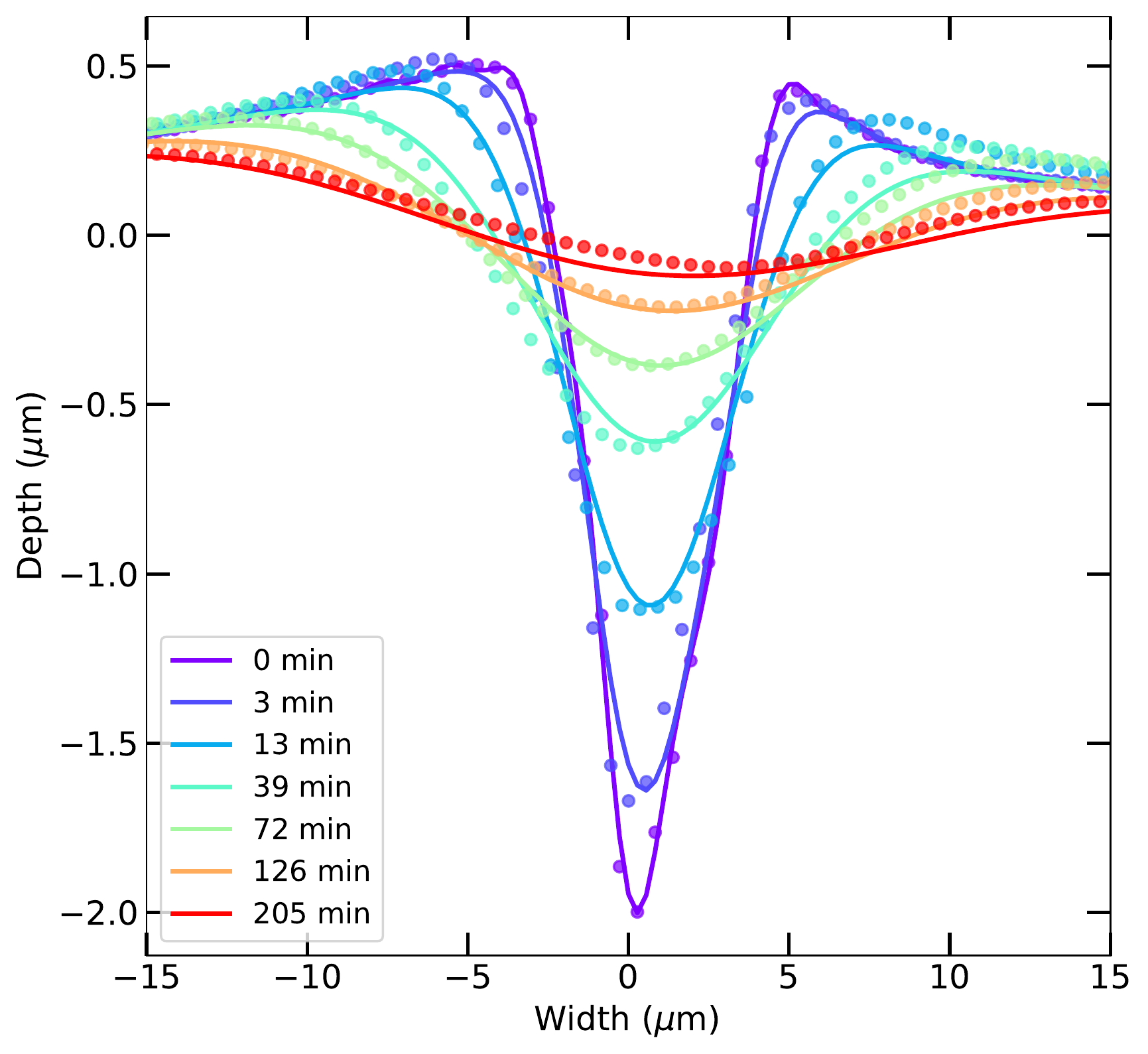}
  \caption{Self-healing of a micron-sized scratch in ice ($T_{\rm ice} = 247$~K). For each time step, dots depict data taken by profilometry, whereas solid lines are fits by the sublimation-condensation model. For clarity, seven time steps are shown of the 22 recorded in total.}
  \label{Fig:ScratchProfilesInTime} 
\end{figure}

Our measurements provide highly detailed images of the ice scratch profile as it slowly self-heals. As illustrated in Fig.\,\ref{Fig:3DPhotographsInTime}, the  initially sharp-edged scratch evolves into a smooth profile, and eventually disappears altogether. To quantify these dynamics and to avoid measuring local impurities in the ice, we average scratch cross-sections over a length of 220 $\mu$m, see Fig.\,\ref{Fig:ScratchProfilesInTime} (solid points). The resulting data allow us to experimentally test the four ice-healing mechanisms proposed thus far: (1) a fluid flow of liquid-like water molecules from the outermost layer; (2) displacement by local sublimation and condensation; (3) movement by volume diffusion as a bulk process; and (4) a rearrangement of the topmost loosely bound molecules by surface diffusion. To this end, we numerically solve the differential equation of each model, and compare the results to the experimentally observed time-dependent scratch profile.

The theoretical basis for each of the four mechanisms was given for the one-dimensional case by Mullins~\cite{mullins1957theory,mullins1959flattening}. The Mullin's model assumes that the attachment and detachment of molecules can occur everywhere on the surface, which is valid for ice with its disordered interface. He derived that in the case of an initial sinusoidal profile with wavelength $\lambda$, only the overall amplitude of the profile changes with time, so that the time-dependent distance of the ice surface with respect to the unscratched surface is given by $U(x,t)=u(t)\sin(2\pi x/\lambda)$, where $x$ is the direction perpendicular to the scratch and $t$ is time. The evolution equation for the amplitude $u(t)$ depends on the mechanism and is given by
\begin{equation}
        \frac{\partial u}{\partial t} = -C_n(T) \left(\frac{2 \pi}{\lambda}\right)^n u,
        \label{Eq:DiffusionEquation}
\end{equation}
where $C_n(T)$ is a temperature-dependent prefactor and $n$ an integer depending on the model: $n=1$ for fluid flow, $n=2$ for sublimation/condensation, $n=3$ for volume diffusion, and $n=4$ for surface diffusion.
Mullins also showed that the equations governing the mass diffusion are linear in the sense that the sum of any two solutions is again a solution~\cite{mullins1959flattening}. Hence, the evolution of an arbitrary initial profile can be obtained from a Fourier analysis, and this is how we calculate the time-dependent profiles for each of the four models: we decompose the initial experimental profile as a Fourier sum (using the 90 lowest-spatial-frequency Fourier components) and propagate each component independently in time. We apply a correction for a small overall slope of the initial profile if necessary. Our analysis involves the following simplifying assumptions: (i) the measurements are carried out in a closed system where the vapor pressures of the flat ice surface and air are in equilibrium
(ii) the mass transfer coefficients are not affected by the crystal orientation of the ice lattice; (iii) the slope of the profiles is small enough to apply the small-slope approximation ($\frac{\partial u}{\partial x} \ll 1$).\\

\begin{figure}
  \centering
  \includegraphics[scale=0.45]{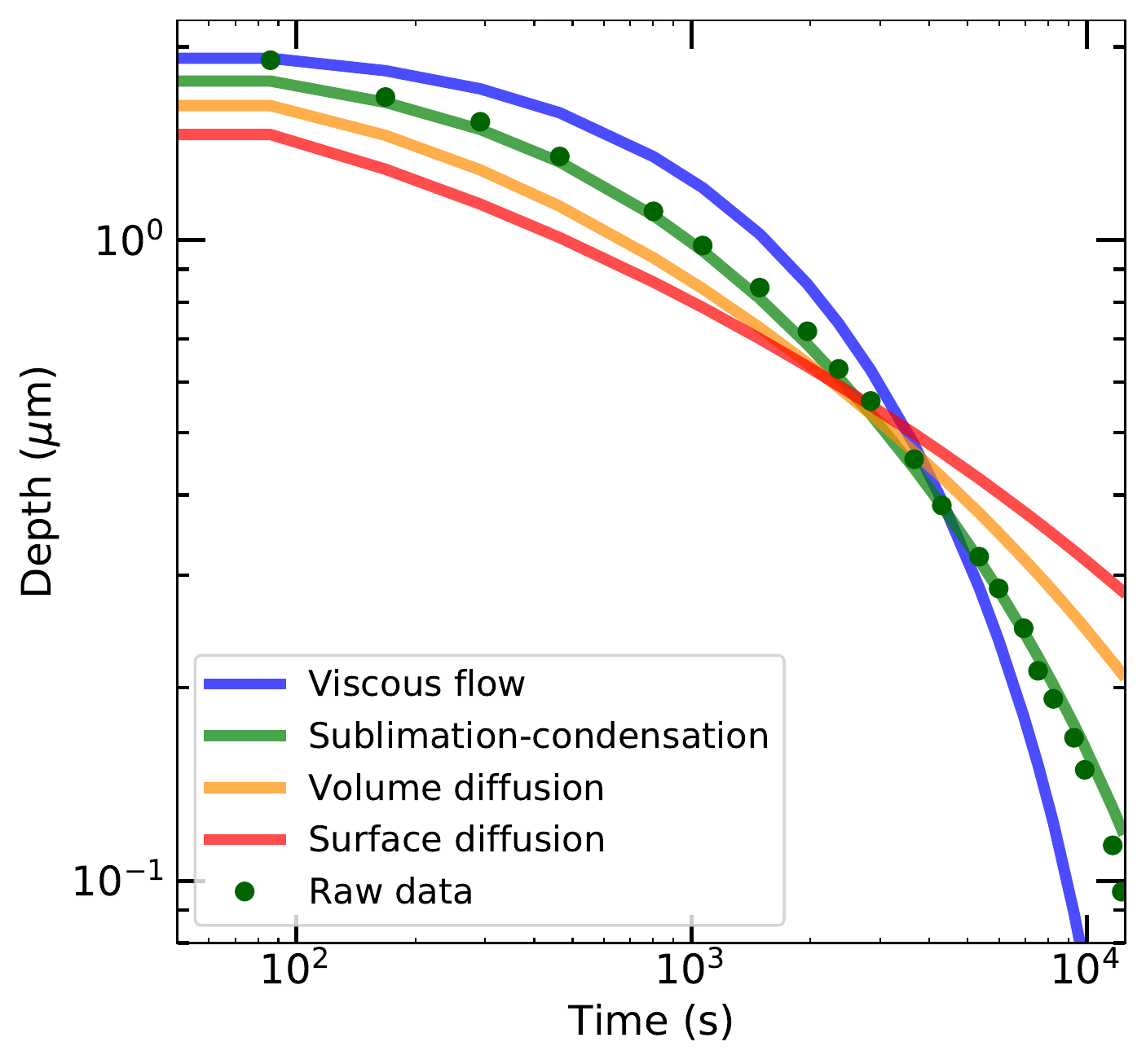}
  \caption{\label{Figure2} Maximum depth of an ice scratch as a function of time. Green dots indicate experimental data; solid lines indicate best fits of the four different candidate models.}
  \label{Fig:DepthinTime}
\end{figure}

We test each mass transfer model by comparing the theoretical prediction with the data of 30 independent measurements using $\chi$-square minimisation, with $C_n(T)$ as the only free parameter. We find that the sublimation-condensation process exhibits the lowest $\chi$-square and thus best agreement with the measurements (Supplementary Fig.~7). 
In Fig.\,\ref{Fig:ScratchProfilesInTime}, we show this agreement for the scratch of Fig.~\ref{Fig:3DPhotographsInTime}, the least-squares fits to the other models are shown in Supplementary Fig.~8. The differences between the four models are further illustrated by plotting the absolute maximum depth of the ice scratch profile developing in time for the best fitting parameters $C_n(T)$ in Fig.\,\ref{Fig:DepthinTime}. Clearly, the best description of the dynamics is given by the sublimation-condensation model.\\ 

In the sublimation-condensation driven healing process, the smoothening of the profile is driven by an increased vapor pressure for curved surfaces (i.e., the Kelvin equation, for the full derivation see Supplementary Information). This self-healing process is relatively fast in the first few minutes and slows down as the surface becomes less curved, as observed in Fig.\,\ref{Fig:DepthinTime}. 

\begin{figure}[t!]
  \centering
  \includegraphics[scale=0.4]{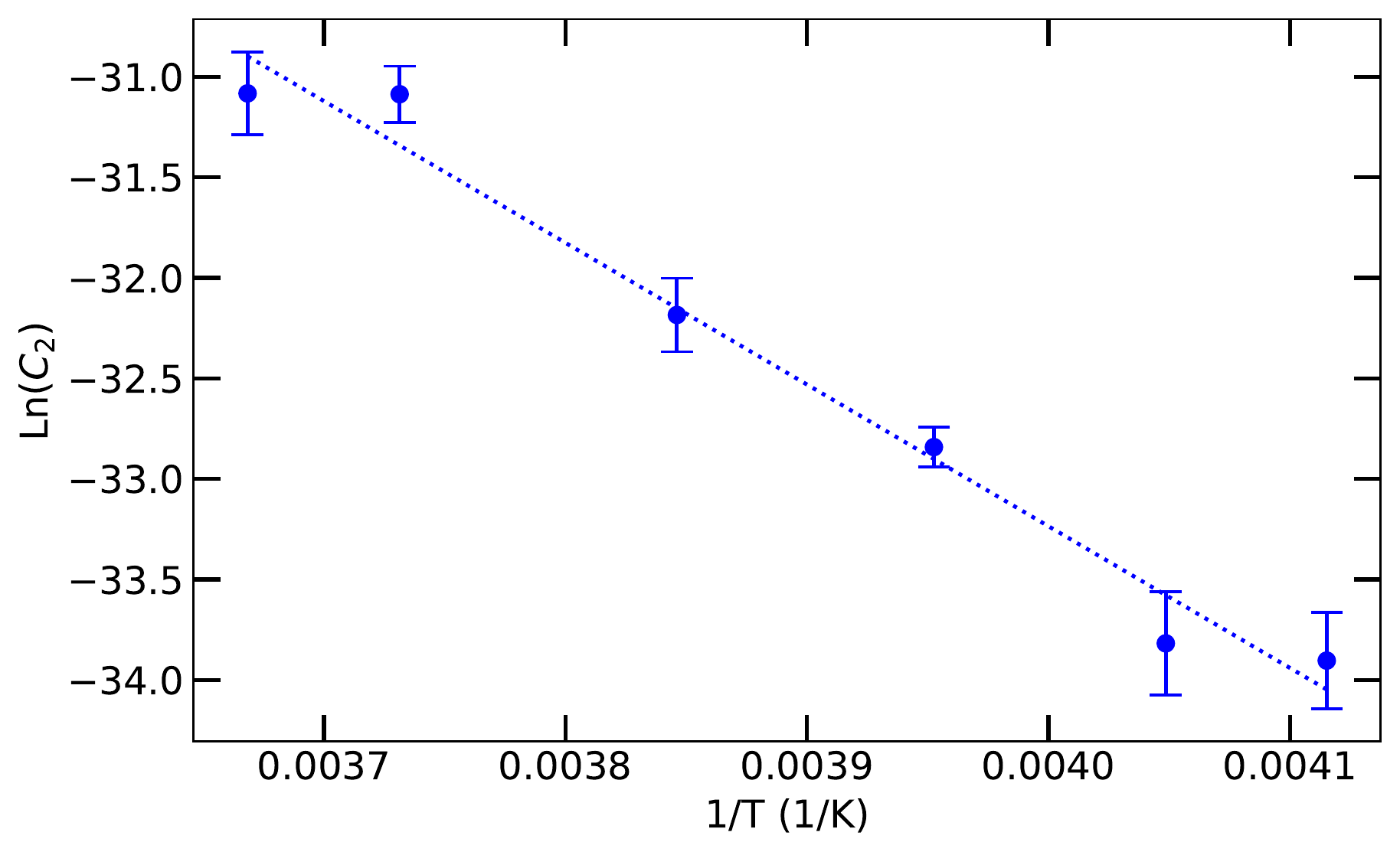}
  \caption{Arrhenius behavior of the sublimation-condensation coefficient $C_2$. Each data point in blue represents 5 measurements on 5 different scratches, a linear fit yields an activation energy of $\Delta E = 58.6 \pm 4.6$ kJ/mole.} 
  \label{Fig:Arrhenius}
\end{figure}

To obtain more insight into the healing mechanism we investigate the temperature dependence of the effective diffusion coefficient. To this end we perform similar least-squares fit analyses for 30 measurements (five scratches profiled at six different temperatures from $243.0$ to $272.6$~K). The effective sublimation-condensation coefficients $C_2(T)$ obtained from the fit follow an Arrhenius temperature dependence with an activation energy of $\Delta E = 58.6 \pm 4.6$ kJ/mole, as shown in Fig.~\ref{Fig:Arrhenius}. 
For comparison, the latent heat for the sublimation of water molecules is approximately $51.1$ kJ/mole \cite{murphy2005review}. The sublimation activation energy of water was found to be in the range $53.1 - 57.3$ kJ/mole \cite{brown1996surface,livingston1998effect,koehler2001desorption}, in good agreement with our result. Moreover, the energy barrier for condensation is significantly lower: $43.35 - 45.1$ kJ/mole\cite{prado2011activation,CRC2005}.   
These results are strong indications that the self-healing of ice is driven by local sublimation instead of local evaporation, before condensation occurs. \\

The above results indicate that the self-healing of ice occurs through a sublimation-condensation mechanism. We now  discuss earlier experimental results that were previously interpreted in terms of the other proposed self-healing mechanisms. 
First, consider the liquid layer interpretation of Faraday. This concept received widespread acclaim by rather precarious comparisons between the physical properties of thin liquid layers and the topmost layer of ice \cite{jellinek1961liquidlike, nakaya1954simple,szabo2007subsecond}. However, 
none of the sintering experiments could be quantitatively reproduced by the liquid layer model~\cite{blackford2007sintering}.

Second, the research that concluded that volumetric bulk diffusion is responsible for the sintering of small ice beads was conducted in a liquid kerosene saturated environment~\cite{kuroiwa1961study}. This inhibited the movement of the molecules along the surface and, more importantly, completely prevented mass diffusion via the vapor phase. Hence, this specific experiment on ice sintering is not generally applicable. 

Third, the research on surface diffusion, done by Kingery, matched the theoretical predictions of the neck-growth between two touching ice spheres. However, the obtained activation energy of approximately $115$ kJ/mole is more than twice the latent heat \cite{kingery1960regelation}. Remarkably, when we force the model for surface diffusion onto the data of our scratch-healing experiments, we obtain a similar activation energy of $\Delta E = 100.0 \pm 11.1$ kJ/mole. 
Weber \textit{et al.} used friction experiments and molecular dynamic simulations on the topmost layer of solid water molecules to demonstrate that the activation energy of surface diffusion is roughly $11.5$ kJ/mole: one order of magnitude lower \cite{weber2018molecular}. 

\begin{figure*}
    \centering
    \begin{subfigure}
        \centering
        \includegraphics[width=0.37\textwidth]{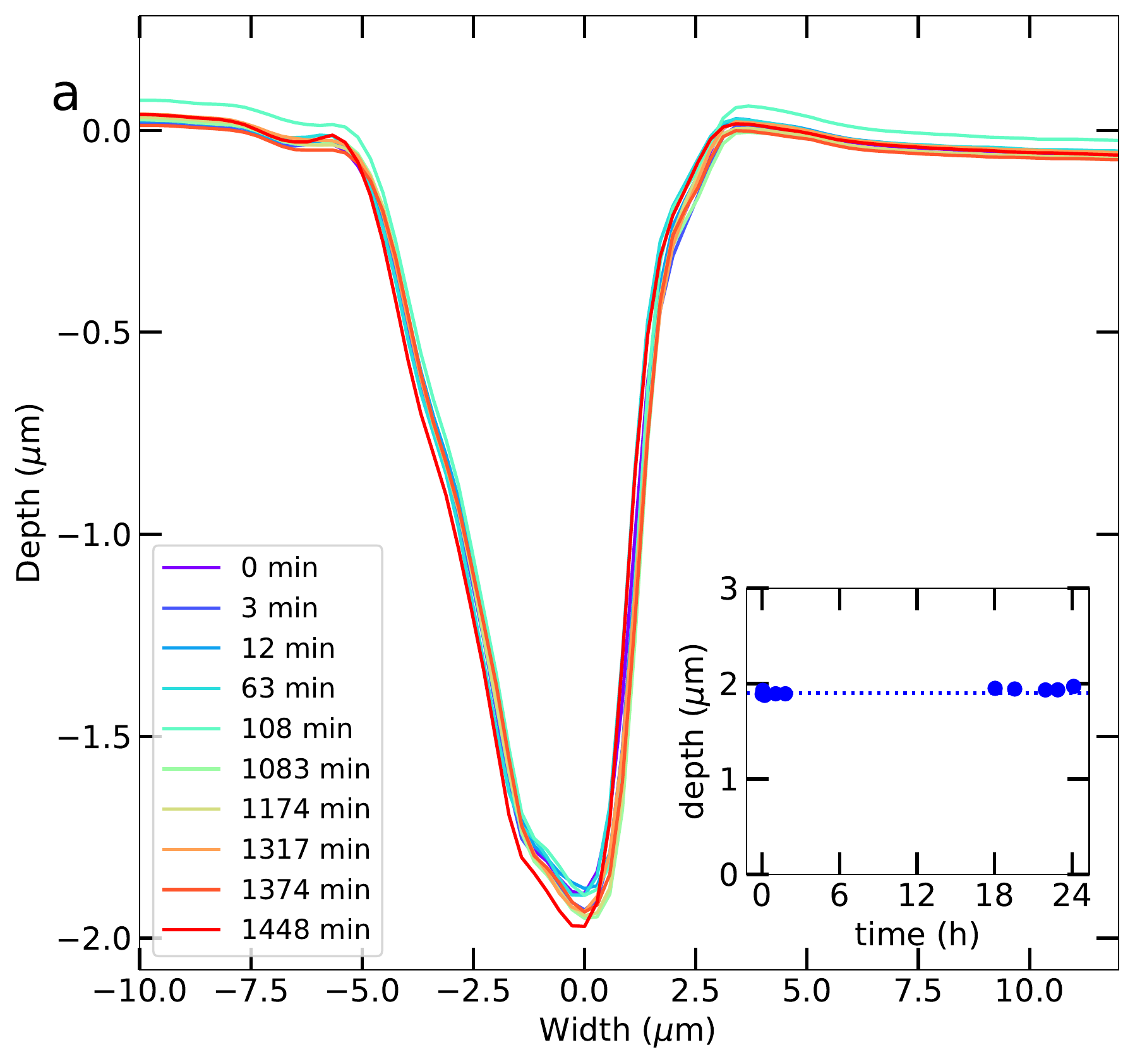}
        \label{Gallium}
    \end{subfigure}
    \begin{subfigure}
        \centering
        \includegraphics[width=0.36\textwidth]{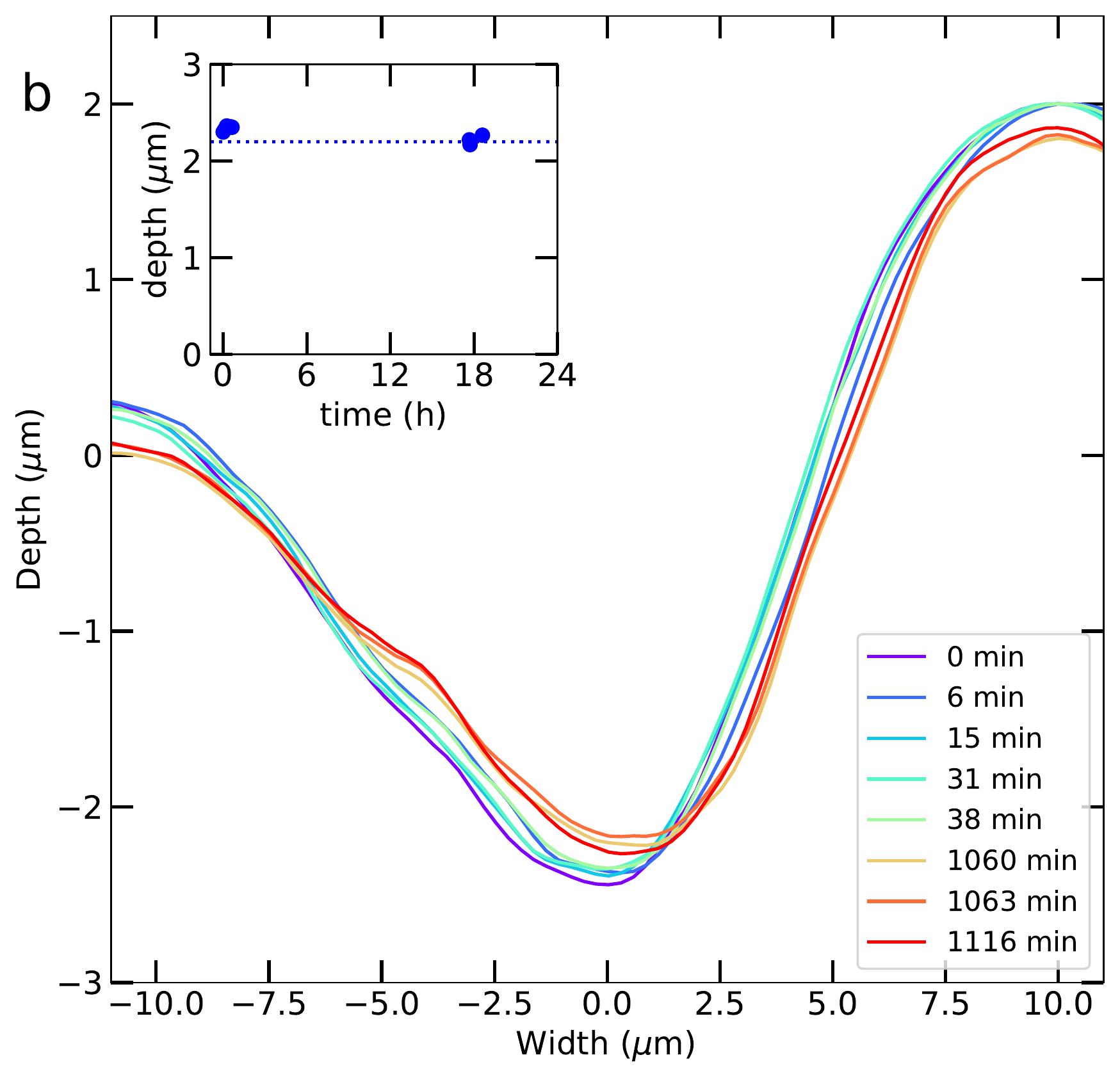}
        \label{CaCl2-6H2O}
    \end{subfigure}
    \caption{Measured profiles of scratched Gallium (a) and $\rm CaCl_2 \cdot 6H_2O$ (b). Both measurements were conducted $8$~K below the melting temperature of the crystalline material, and do not show evidence of self-healing. The insets depict the time-evolution of the absolute depth. The height difference in the profile of $\rm CaCl_2 \cdot 6H_2O$ is caused by the plowing effect while scratching.}
    \label{fig:gallium and CaCl2-6H2O}
\end{figure*}

To investigate to what extent the self-healing behavior is unique to ice, we also study the evolution of scratches in other crystalline materials. In particular, we measured the metal gallium and the mineral $\rm CaCl_2 \cdot 6H_2O$, both $8$~K below their melting temperature (information about the preparation of these samples can be found in the SI). In both materials, surface diffusion is expected to be mainly responsible for the rate-limiting process because of their lower vapor pressures. Hence, micron-sized scratches are not seen to self heal in an experimentally accessible time scale (Fig.~\ref{fig:gallium and CaCl2-6H2O}). Compared to these, and other solid materials, ice is unique in its high vapor pressure~\cite{murphy2005review,marti1993survey}. Previous studies have shown, in agreement with our findings, that this causes sublimation with an increased rate at high surface curvatures~\cite{jambon2018singular}. Also note that, $8$~K below the melting temperature, the poly-crystalline gallium is in the premelting regime, where the surface is in a quasi-liquid state~\cite{danzig2021surface}. This emphasizes once more that self-healing cannot simply be explained by the existence of a quasi-liquid layer. 

To conclude, we find that the detachment and reattachment of highly mobile water molecules on the ice surface causes scratches in the ice surface to self-heal spontaneously. By quantitatively studying the self-healing behavior of micrometer-sized scratches, and comparing the results with four models proposed for the transport of molecules on the ice surface, we conclude that the main mechanism of transport is through sublimation and condensation.
We propose that the efficient self healing of ice compared to other materials might be due to the water molecules in ice being connected by hydrogen bonds: in contrast to the attractive interactions in the crystals of most other materials, hydrogen bonding is highly cooperative, meaning that breaking four hydrogen bonds in the bulk requires much more than two times the energy required for breaking two hydrogen bonds at the interface. As a consequence, the water molecules at the surface can detach relatively easily, even though the bulk crystal phase is completely stable. 

\begin{acknowledgement}

This work is financially supported by an NWO Projectruimte grant \#680-91-133.

\end{acknowledgement}
\section*{Supplementary information}
\noindent For the derivation of sublimation-condensation model (n = 2), we start with the Kelvin equation, showing the relation between the vapor pressure of a surface with respect to a curved surface with radius $r$:
\begin{align}
    {\rm ln} \left( \frac{P}{P_{sat}} \right) =\frac{2\gamma V_m}{rRT} ,
\end{align}
\noindent where $P$ and $P_{sat}$ are the vapor pressures of the curved and flat surface, respectively; $\gamma$ is the surface tension, $V_m$ the molar volume, $R$ the universal gas constant, and $T$ the temperature. Using the definition of curvature $\kappa = 1/r$, we can define the Kelvin equation along the curve of the scratch:
\begin{align}
    {\rm ln} \left( \frac{P}{P_{sat}} \right) =\frac{2\kappa\gamma V_m}{RT}
\end{align}
Since we are measuring relatively small curvatures we can use the first order Taylor expansion for $ln(P/P_{sat})$ around ln(1) = $(P/P_{sat} - 1)$ and call $P-P_{sat} = \Delta P$. Then we obtain:
\begin{align}
    \frac{\Delta P}{P_{sat}}=\frac{2\kappa\gamma V_m}{RT}
\end{align}
\noindent Classic kinetic theory of gasses states that the number of emitted atoms per surface area per second $\Delta\theta$ is given by:
\begin{align}
    \Delta \theta = \frac{\Delta P}{\sqrt{2 \pi M RT}},
\end{align}
with $M$ the molecular weight. 
Note that we omit an extra Arrhenius factor on the right-hand side, that usually represents the attachment or detachment process in crystal growth theory. This term is negligible in the case of ice crystals since the limiting process is the diffusion of water molecules in the vapor phase.
Filling in $\Delta P$ from Eq.\,4, and multiplying both sides with $V_m$ yields the velocity of the volume per unit area of emitted particles:
\begin{align}
    V_m\Delta \theta = \sqrt{\frac{2}{\pi M}}\frac{P_{sat}\kappa \gamma V_m^2}{(RT)^\frac{3}{2}}
\end{align}
The curvature $\kappa$ can be denoted as:
\begin{align}
    \kappa = {\frac{\partial^2 y}{\partial x^2}}{\left(1+\left( \frac{\partial y}{\partial x}\right) ^2\right)^{-\frac{3}{2}}},
\end{align}
and the projected velocity in the $y$-direction of $V_m \Delta \theta$ as:
\begin{align}
    V_m \Delta \theta = \sqrt{1+\left(\frac{\partial y}{\partial x}\right)^2}\frac{\partial y}{\partial t}
\end{align}
Together with the small slope approximation $\frac{\partial y}{\partial x} \ll1$, we obtain:
\begin{align}
    \frac{\partial y(x,t)}{\partial t} = C_2(T)\frac{\partial^2 y(x,t)}{\partial x^2},
\end{align}
which results, for a sinusoidal case, in:
\begin{align}
        \frac{\partial y(x,t)}{\partial t} = -C_2(T) k^2 y(x,t),
\end{align}
with $C_2(T) = \sqrt{2}P_{sat}\gamma V_m^2/((\pi M)^\frac{1}{2}(RT)^\frac{2}{3})$ and $k = \frac{2 \pi}{\lambda}$.
\\\\

\newpage

\begin{figure}[h!]
  \centering
  \includegraphics[scale=0.25]{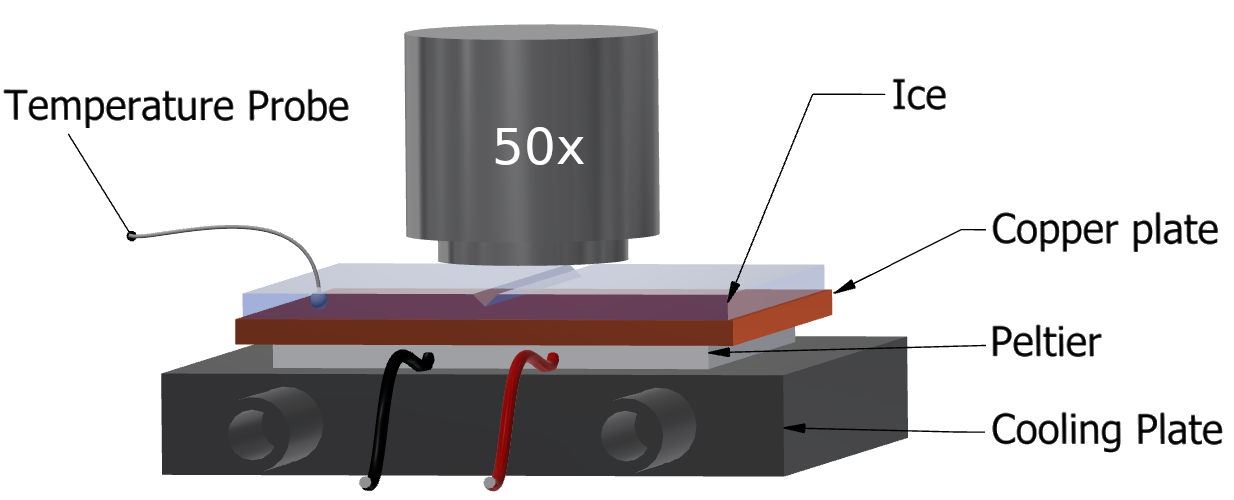}
  \caption{\label{Figureminus1} Schematic view of the part of the experimental setup that was placed on a motorized stage with sub-micron lateral precision. This system was placed in a temperature and humidity controlled chamber.}
\end{figure}

\begin{figure}[h]
  \centering
  \includegraphics[scale=0.4]{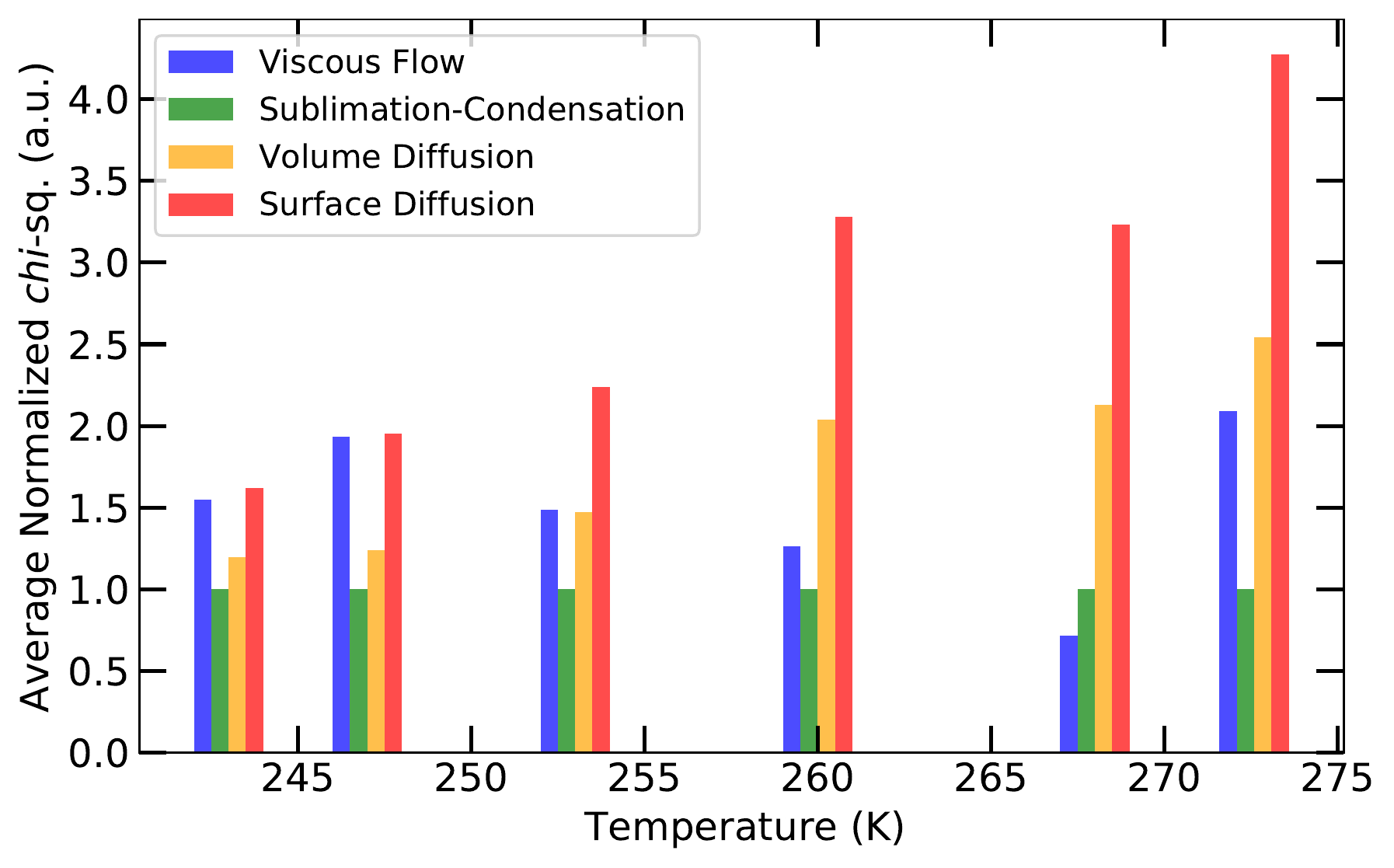}
  \caption{\label{Figure4} $\chi$-square quantification of the best fits of each of the 4 models to the experimental data at 6 temperatures. For each temperature we averaged 5 measurements that were normalized to the $\chi$-square value of the sublimation-condensation model.}
\end{figure}

\begin{figure*}[h]
    \centering
    \begin{subfigure}
        \centering
        \includegraphics[width=0.36\textwidth]{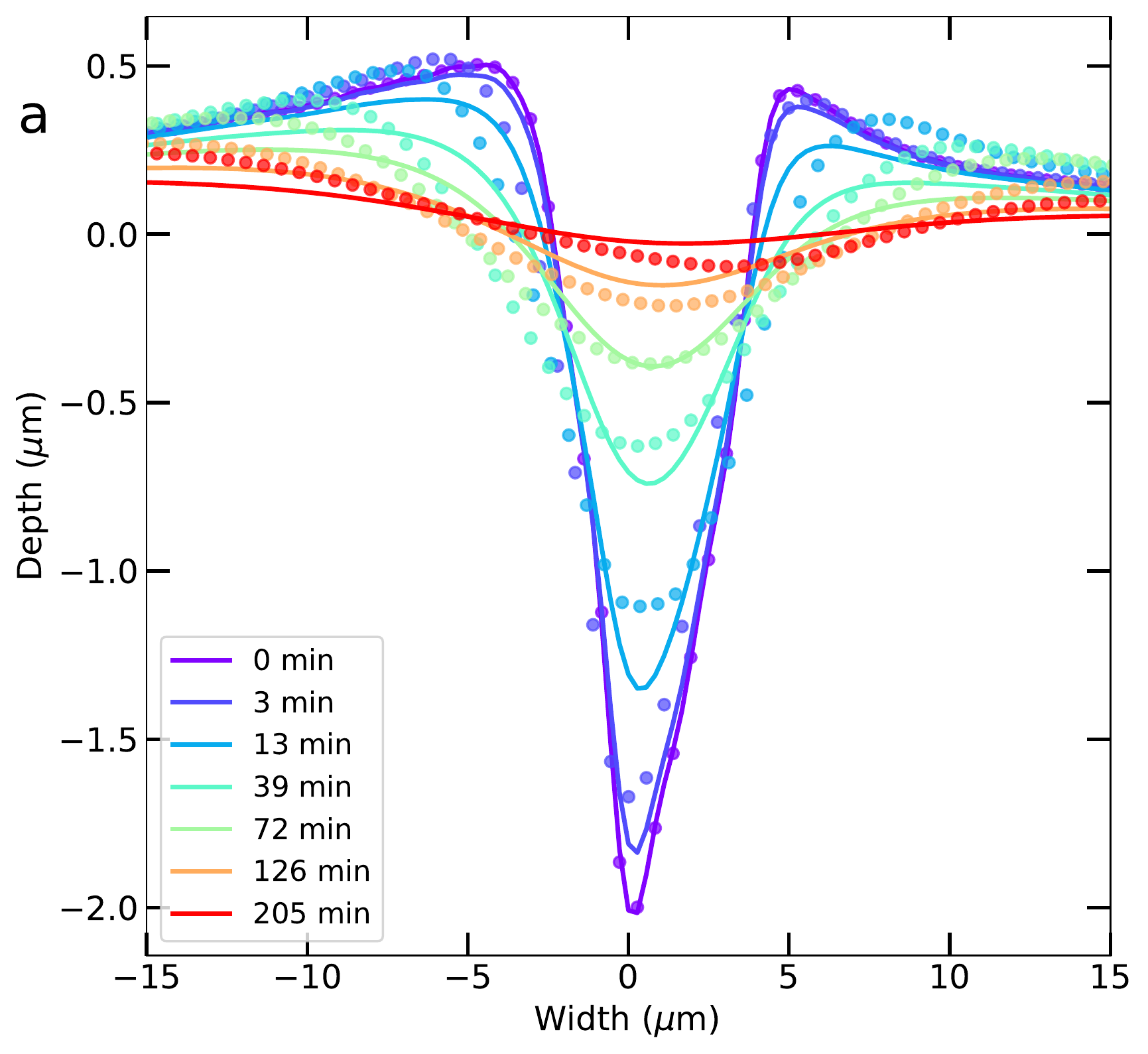}
        \label{1st order}
    \end{subfigure}
    \begin{subfigure}
        \centering
        \includegraphics[width=0.36\textwidth]{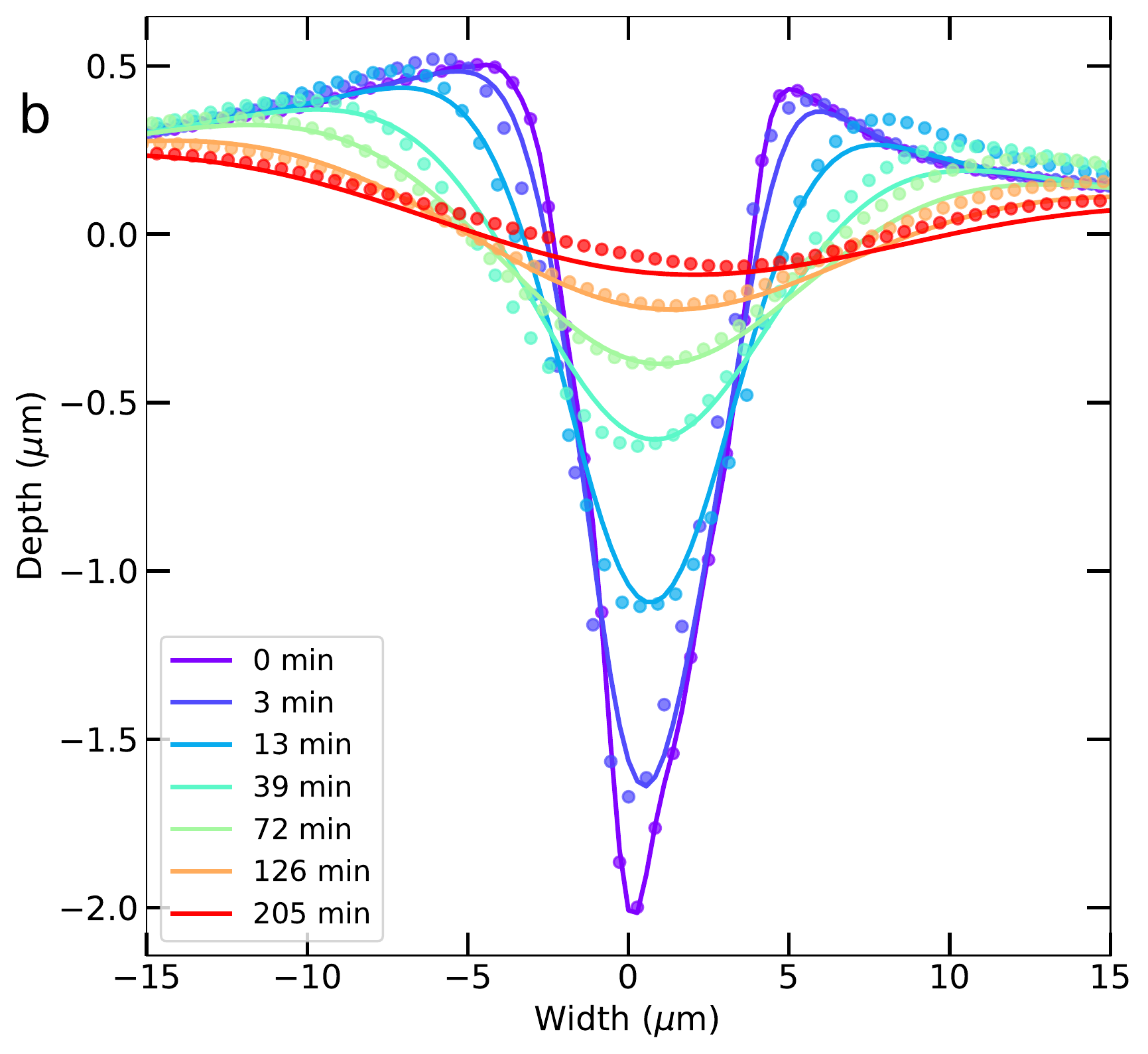}
        \label{2nd order}
    \end{subfigure}
    \\
    \centering
    \begin{subfigure}
        \centering
        \includegraphics[width=0.36\textwidth]{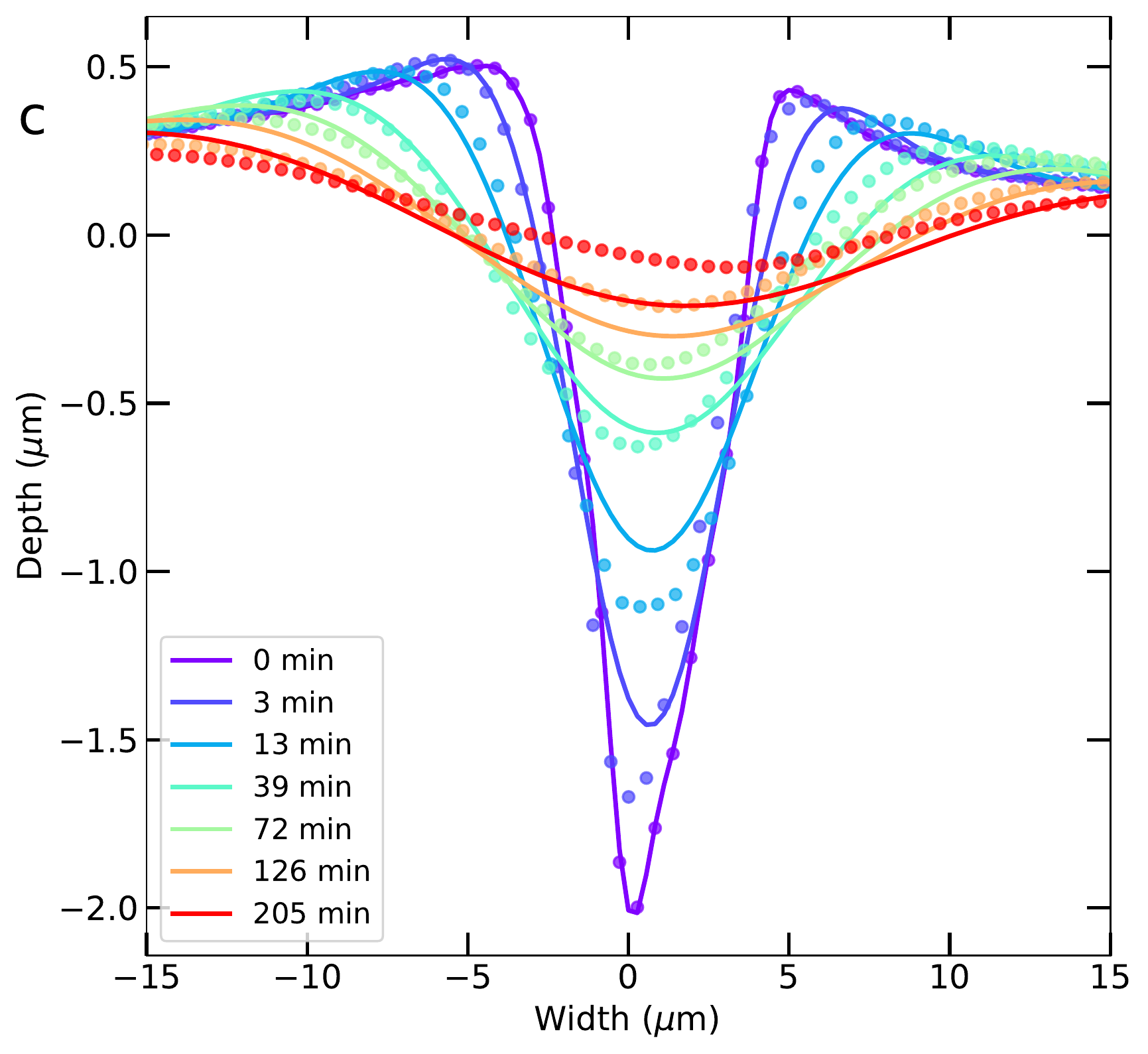}
        \label{3rd order}
    \end{subfigure}
    \begin{subfigure}
        \centering
        \includegraphics[width=0.36\textwidth]{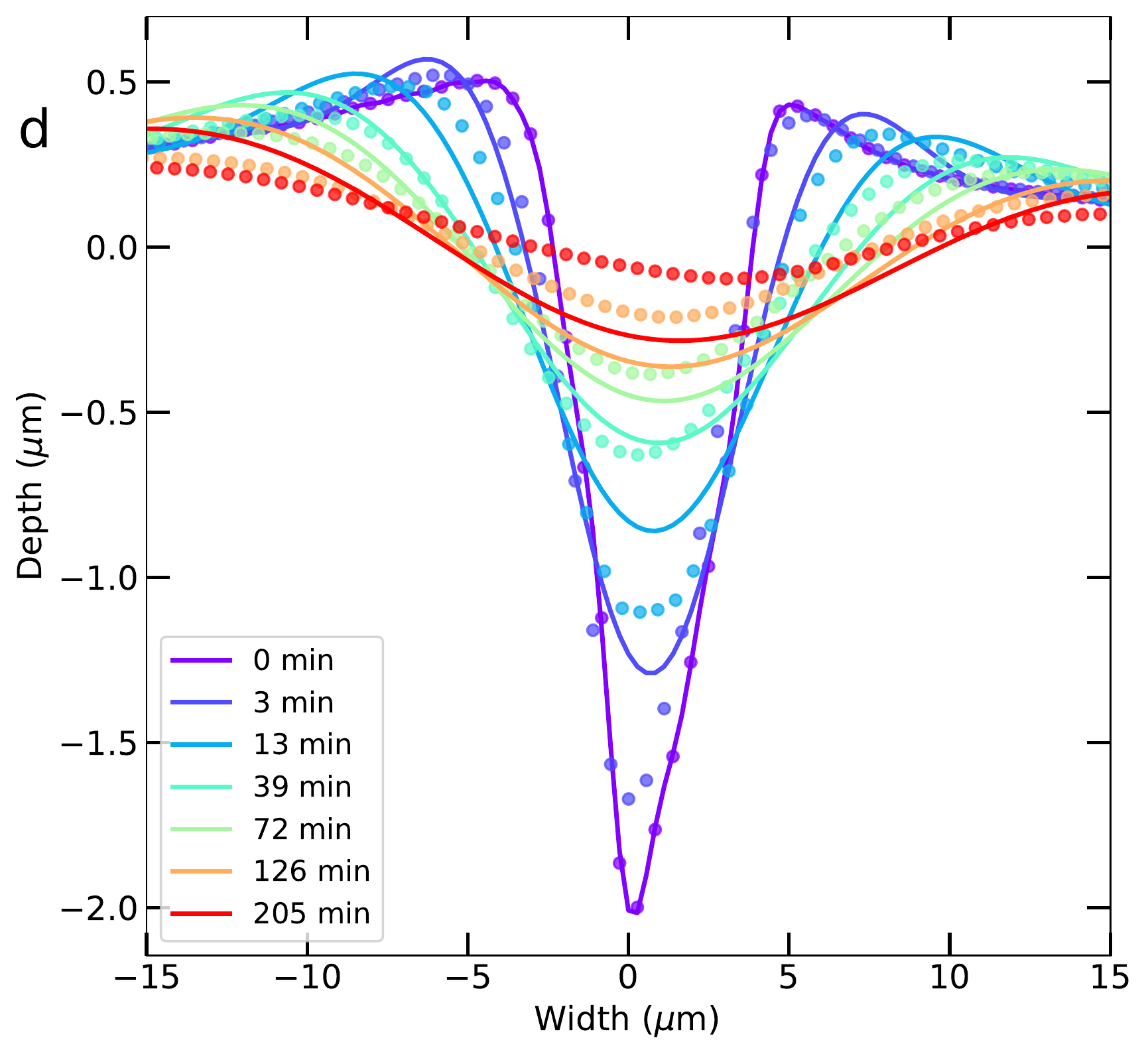}
        \label{4rth order}
    \end{subfigure}
    \caption{Comparison between the four models for the self-healing of a micron-sized scratch in ice ($T_{\rm ice} = 247$~K). For each time step, dots depict data taken by profilometry, whereas solid lines are fits by the viscous flow (a), sublimation and condensation (b), volumetric bulk diffusion (c) and surface diffusion (d) model. For clarity, seven time steps are shown of the 22 recorded.}
    \label{fig:abcd}
\end{figure*}

\newpage
\subsection*{Additional information on the preparation of $\rm CaCl_2 \cdot 6H_2O$ and gallium}
\noindent $\rm CaCl_2 \cdot 6H_2O$ polycrystals (melting temperature of appoximately 30 °C) have been prepared by dissolving anhydrous $\rm CaCl_2$ powder (Merck) in Milli-Q-water until the 90 \% saturation had been reached (Solubility in water = 74.5 g/100 mL at 20 °C). The salt solution was then poured into a petri dish and stored in a conditioned environment with T = 21 °C and at relative humidity RH = 4.5 \%. The $\rm CaCl_2 \cdot 6H_2O$ crystallization occurred through evaporation of the water in time. 

To create a pristine gallium surface (99,997\% purity, commercially available at novaelements), we first heated the metal above its melting point  (29.76 °C) in a petri dish. Crystallization was induced by cooling down the gallium to room temperature and by adding a tiny solid piece of solid gallium. 


\bibliography{achemso-demo}
\end{document}